\documentclass[preprint]{elsarticle}
\usepackage{amsmath,amsfonts,amssymb,bm}
\numberwithin{equation}{section}
\newcommand{\abs}[1]{\left\vert#1\right\vert}
\DeclareMathOperator{\erf}{erf}
\begin{document}
\title{Path integrals of systems on the half-line}
\author{Seiji Sakoda}
\ead{sakoda@nda.ac.jp}
\address{Department of Applied Physics, National Defense Academy,
Hashirimizu\\
Yokosuka city, Kanagawa 239-8686, Japan}

\date{\today}

\begin{abstract}
It is wellknown that the Feynman kernel for the free particle on the half-line can be expressed as a sum over classical paths if we take the contribution from the reflected path into account. The minus sign for the reflected path needs to be generalized depending on the parameter of the potential if we add a potential proportional to $1/x^{2}$.
\end{abstract}
\maketitle

\section{Introduction}
The Feynman kernel for the free particle on the half-line\cite{SchulmanBook,KleinertBook} $0<x<\infty$ can be expressed as a sum of contributions from the path which connects the endpoint with the starting one directly and the path reflected once at the origin. The direct path yields
\begin{equation}
\label{eq:amp01}
	\sqrt{\frac{m}{\ 2\pi\hbar\beta\ }\,}
	\exp\left\{-\frac{m}{\ 2\hbar\beta\ }(a-b)^{2}\right\}
\end{equation}
while the one from the reflected path being given by
\begin{equation}
\label{eq:amp02}
	-\sqrt{\frac{m}{\ 2\pi\hbar\beta\ }\,}
	\exp\left\{-\frac{m}{\ 2\hbar\beta\ }(a+b)^{2}\right\}
\end{equation}
as contributions to the Feynman kernel with imaginary time $\beta$.
Here the factor $-1$ for the reflected path makes the sum of contributions above vanish at the origin if we set $a$ or $b$ to be equal to $0$. Therefore the Feynman kernel 
\begin{equation}
\label{eq:kernel01}
	K(a,b;\beta)=\sqrt{\frac{m}{\ 2\pi\hbar\beta\ }\,}\left[
	\exp\left\{-\frac{m}{\ 2\hbar\beta\ }(a-b)^{2}\right\}-
	\exp\left\{-\frac{m}{\ 2\hbar\beta\ }(a+b)^{2}\right\}\right]
\end{equation}
fulfills the boundary condition $K(0,b;\beta)=K(a,0;\beta)=0$ in addition to the imaginary time version of the Schr\"{o}dinger equation:
\begin{equation}
\label{eq:schrodinger}
	-\hbar\frac{\ \partial\ \ }{\ \partial\beta\ }K(a,b;\beta)=
	-\frac{\ \hbar^{2}\ }{\ 2m\ }\frac{\ \partial^{2}\ }{\ \partial a^{2}\ }
	K(a,b;\beta).
\end{equation}
By making use of the Fourier transform, we can convert the kernel above into
\begin{equation}
\label{eq:kernel02}
	K(a,b;\beta)=\int_{0}^{\infty}\!\!
	e^{-\lambda k^{2}/2}\phi_{k}(a)\phi_{k}(b)\,dk,\quad
	\lambda\equiv\frac{\ \hbar\beta\ }{m},
\end{equation}
where the eigenfunction of the free particle on the half-line being given
$\phi_{k}(x)=\sqrt{2/\pi\,}\sin(kx)$ for the energy eigenvalue $E_{k}=(\hbar k)^{2}/(2m)$.

In this article, we formulate and consider decomposition into a sum over paths of the Feynman kernel for a particle whose Hamiltonian being given by
\begin{equation}
\label{eq:ham01}
	H^{(\nu)}=\frac{\ p^{2}\ }{\ 2m\ }+
	\frac{\ \nu(\nu-1)\hbar^{2}\ }{\ 2mx^{2}\ }+V(x),
\end{equation}
with the assumption $\nu\ge1/2$, on the half-line $0<x<\infty$. Here, in the Hamiltonian above, we have introduced an additional potential $V(x)$ which may be chosen to be proportional to $x^{2}$ for the radial oscillator and to $1/x$ for the radial Coulomb problem. If we set $\nu=1$ as well as $V(x)=0$ above, the Hamiltonian reduces to the one for the free particle on the half-line. Therefore $H^{(\nu)}$ is a generalization to the free particle. As we shall see below, this generalization results in a phase factor $e^{\nu\pi i}$ instead of $-1$ for the contribution by the reflected path to the Feynman kernel for the free particle.

\section{Formulation of the path integral}
To formulate a path integral for the Hamiltonian $H^{(\nu)}$ defined in the previous section, we first consider the stationary Schr\"{o}dinger equation
\begin{equation}
\label{eq:schrodinger01}
	\left\{-\frac{\ d^{2}\ \ }{\ dx^{2}\ }+
	\frac{\ \nu(\nu-1)\ }{x^{2}}\right\}\phi_{k}^{(\nu)}(x)=
	k^{2}\phi_{k}^{(\nu)}(x)
\end{equation}
of the Hamiltonian
\begin{equation}
\label{eq:ham02}
	H_{0}^{(\nu)}=\frac{\ p^{2}\ }{\ 2m\ }+
	\frac{\ \nu(\nu-1)\hbar^{2}\ }{\ 2mx^{2}\ }
\end{equation}
for an eigenvalue $E_{k}=(\hbar k)^{2}/(2m)$ ($k\ge0$).
The normalized solution for \eqref{eq:schrodinger01} is given in terms of the Bessel function by
\begin{equation}
\label{eq:sol01}
	\phi_{k}^{(\nu)}(x)=\sqrt{kx\,}J_{\nu-1/2}(kx).
\end{equation}
On this eigenfunction, there holds
\begin{equation}
\label{eq:orth}
	\int_{0}^{\infty}\!\!\phi_{k}^{(\nu)}(x)\phi_{k{'}}^{(\nu)}(x)\,dx=
	\delta(k-k{'})
\end{equation}
as the orthogonality relation for $k>0$ and $k{'}>0$ in addition to the completeness
\begin{equation}
\label{eq:comp}
	\int_{0}^{\infty}\!\!\phi_{k}^{(\nu)}(x)\phi_{k}^{(\nu)}(x{'})\,dk=
	\delta(x-x{'})
\end{equation}
for $x>0$ and $x{'}>0$.
Note that these are understood as the limit ($a\to0$) in a formula for positive $p$ and $q$
\begin{equation}
\label{eq:formula}
	\int_{0}^{\infty}\!\!xe^{-a^{2}x^{2}}
	J_{\mu}(px)J_{\mu}(qx)\,dx=
	\frac{1}{\ 2a^{2}\ }e^{-(p^{2}+q^{2})/(4a^{2})}
	I_{\mu}\left(\frac{\ pq\ }{\ 2a^{2}\ }\right)
\end{equation}
in which $\Re(\mu)>-1$ as well as $\abs{\arg(a)}<\pi/4$ being assumed. We make use of the completeness above though there exists another method to obtain path integrals on the half-line\cite{Fujikawa2008,Sakoda2017} in this article. On earlier attempts on the path integral on the half-line or the path integral in radial coordinate, see Chap. 8 of Ref.\cite{KleinertBook} and references therein.

In the asymptotic expansion of the modified Bessel function, we usually pick up the leading term and drop second term, that is exponentially small if $\Re(z)$ is positive, in
\begin{equation}
\label{eq:asymp01}
	I_{\nu}(z)\sim\frac{1}{\ \sqrt{2\pi z\,}\ }e^{z}
	\sum_{k=0}^{\infty}(-1)^{k}\frac{\ (\nu,k)\ }{\ (2z)^{k}\ }+
	\frac{\ e^{-z+(\nu+1/2)\pi i}\ }{\sqrt{2\pi z\,}}
	\sum_{k=0}^{\infty}\frac{\ (\nu,k)\ }{\ (2z)^{k}\ },
\end{equation}
where $(\nu,k)$ is defined by 
\begin{equation}
\label{eq:asymp02}
	(\nu,k)=\frac{\ \varGamma(\nu+k+1/2)\ }{\ k!\varGamma(\nu-k+1/2)\ }
\end{equation}
and $-\pi/2<\arg(z)<3\pi/2$ has been assumed for $\abs{z}\to\infty$.
If we retain the second term above, we will obtain
\begin{equation}
\label{eq:comp01}
	\int_{0}^{\infty}\!\!\phi_{k}^{(\nu)}(x)\phi_{k}^{(\nu)}(x{'})\,dk=
	\delta(x-x{'})+e^{\nu\pi i}\delta(x+x{'})
\end{equation}
instead of \eqref{eq:comp}.

Although the second delta function in the right hand side of \eqref{eq:comp01} has no support for positive $x$ and $x{'}$, in view of the Feynman kernel for the free particle on the half-line, we should retain this term to take into account of the contribution by the reflected path in considering the path integral on the half-line. Keeping this in mind, we examine the short time kernel
\begin{equation}
\label{eq:kernel03}
	K(x,x{'};\epsilon)=\langle x\vert
	\left(1-\frac{\ \epsilon H^{(\nu)}\ }{\hbar}\right)
	\vert x{'}\rangle
\end{equation}
for infinitesimally small $\epsilon$. For $V(x)$ in the Hamiltonian, we can evaluate it as
\begin{equation}
\label{eq:pote01}
	\langle x\vert V(x)\vert x{'}\rangle=V(\sqrt{xx{'}\,})
	\langle x\vert x{'}\rangle.
\end{equation}
Then, by making use of the completeness \eqref{eq:comp}, we obtain
\begin{equation}
\label{eq:kernel04}
	K(x,x{'};\epsilon)=\sqrt{xx{'}\,}\int_{0}^{\infty}\!\!kdk\,
	J_{\nu-1/2}(kx)J_{\nu-1/2}(kx{'})\left[
	1-\frac{\ \epsilon\ }{\ \hbar\ }\left\{
	\frac{\ (\hbar k)^{2}\ }{\ 2m\ }+V(\sqrt{xx{'}\,})\right\}\right].
\end{equation}
After exponentiating the expression in the bracket above, we carry out the integration with respect to $k$ to find
\begin{equation}
\label{eq:kernel05}
	K(x,x{'};\epsilon)=\frac{\ \sqrt{xx{'}\,}\ }{\ \lambda\ }
	I_{\nu-1/2}\left(\frac{\ xx{'}\ }{\lambda}\right)
	\exp\left\{-\frac{1}{\ 2\lambda\ }(x^{2}+x{'}{}^{2})
	-\frac{\ \epsilon\ }{\ \hbar\ }V(\sqrt{xx{'}\,})\right\},
\end{equation}
where an abbreviation is introduced to write $\hbar\epsilon/m$ as $\lambda$. 
We set $\epsilon=\beta/N$ for large $N$ here to obtain the Feynman kernel
\begin{equation}
\label{eq:feynman01}
	K(a,b;\beta)=\lim\limits_{N\to\infty}\int_{0}^{\infty}
	\prod_{j=1}^{N-1}dx_{j}\,
	K(a,x_{N-1};\epsilon)K(x_{N-1},x_{N-2};\epsilon)\times\cdots\times
	K(x_{1},b;\epsilon)
\end{equation}
for the Hamiltonian $H^{(\nu)}$. In this way, we obtain a time sliced path integral for a particle on the half-line. 

\section{Evaluation of the path integral}
If we set $\nu=1$ as well as $V(x)=0$ to make use of $I_{1/2}(z)=\sqrt{2/(\pi z)\,}\sinh z$, we find the path integral for a free particle to be given by
\begin{equation}
\label{eq:feynman02}
\begin{aligned}
	K(a,b;\beta)=&\lim\limits_{N\to\infty}
	\left(\frac{1}{\ 2\pi\lambda\ }\right)^{N/2}\int_{0}^{\infty}
	\prod_{j=1}^{N-1}dx_{j}\\
	&\times
	\prod_{k=1}^{N}\left[
	\exp\left\{-\frac{1}{\ 2\lambda\ }(x_{k}-x_{k-1})^{2}\right\}-
	\exp\left\{-\frac{1}{\ 2\lambda\ }(x_{k}+x_{k-1})^{2}\right\}
	\right],
\end{aligned}
\end{equation}
where we have set $x_{N}=a$, $x_{0}=b$ and $\lambda=\hbar\epsilon/m$.
Since the short time kernel is already exact for the free particle, the time sliced path integral above is carried out to result in
\begin{equation}
\label{eq:feynman-3}
	K(a,b;\beta)=\sqrt{\frac{m}{\ 2\pi\hbar\beta\ }\,}
	\left[
	\exp\left\{-\frac{m}{\ 2\hbar\beta\ }(a-b)^{2}\right\}-
	\exp\left\{-\frac{m}{\ 2\hbar\beta\ }(a+b)^{2}\right\}
	\right].
\end{equation}

Let us now consider the short time kernel for $\nu=2$ while keeping $V(x)=0$. The explicit form of the short time kernel is given by
\begin{equation}
\label{eq:feynman03}
	K(x,x{'};\epsilon)=\frac{\ \sqrt{xx{'}\,}\ }{\ \lambda\ }
	I_{3/2}\left(\frac{\ xx{'}\ }{\ \lambda\ }\right)
	\exp\left\{-\frac{1}{\ 2\lambda\ }(x^{2}+x{'}{}^{2})\right\}.
\end{equation}
Since the expression for this kernel is already exact, we can easily validate the reproducing property of the kernel:
\begin{equation}
\label{eq:reproduce}
	\int_{0}^{\infty}\!\!K(x,y;\epsilon)K(y,x{'};\epsilon)\,dy=
	K(x,x{'};2\epsilon)
\end{equation}
by making use of a formula
\begin{equation}
	\int_{0}^{\infty}\!\!xe^{-a^{2}x^{2}}I_{\mu}(px)I_{\mu}(qx)\,dx=
	\frac{1}{\ 2a^{2}\ }e^{(p^{2}+q^{2})/(4a^{2})}
	I_{\mu}\left(\frac{\ pq\ }{\ 2a^{2}\ }\right).
\end{equation}
On the other hand, we can rewrite \eqref{eq:feynman03}, if we recall $I_{3/2}(z)=\sqrt{2/(\pi z)\,}(\cosh z-z^{-1}\sinh z)$, as
\begin{equation}
\label{eq:feynman04}
\begin{aligned}
	K(x,x{'};\epsilon)=&\frac{1}{\ \sqrt{2\pi\lambda\,}\ }
	\left[\exp\left\{-\frac{1}{\ 2\lambda\ }(x-x{'})^{2}-
	\frac{\ \lambda\ }{\ xx{'}\ }\right\}\right.\\
	&\hphantom{\frac{1}{\ \sqrt{2\pi\lambda\,}\ }}\!\!\!
	\left.+
	\exp\left\{-\frac{1}{\ 2\lambda\ }(x+x{'})^{2}+
	\frac{\ \lambda\ }{\ xx{'}\ }\right\}\right]
\end{aligned}
\end{equation}
by keeping only relevant terms in the exponent.
Clearly, the second term in the bracket above expresses the contribution from the reflected path and has its support only on $-\infty<x<0$ if $x{'}>0$ while the first term having its support only on $0<x<\infty$ under the same condition. By this restriction, both vanish at the origin by themselves. Namely, though the boundary condition was satisfied by the cancellation in contributions from direct and reflected paths for the kernel of the free particle, each path contribute in an additive manner and the vanishing of the kernel at the origin is realized by the vanishing property of each component for the case of $\nu=2$.
We can repeat the same procedure for positive integer values of $\nu$ to find that the coefficient of the contribution by the reflected path becomes $-1$ if $\nu$ is an odd integer while the one being given by $+1$ if $\nu$ is an even integer.

We show here an interesting calculation on an integral of a product of the kernel \eqref{eq:feynman04}. In the product $K(x,y;\epsilon)K(y,x{'};\epsilon)$, we have four terms
\begin{equation}
\label{eq:product}
\begin{aligned}
	&\exp\left[-\frac{1}{\ 2\lambda\ }\left\{
	(x-y)^{2}+(y-x{'})^{2}\right\}-
	\frac{\ \lambda\ }{\ y\ }
	\left(\frac{\ 1\ }{\ x\ }+\frac{\ 1\ }{\ x{'}\ }\right)\right]\\
	+&
	\exp\left[-\frac{1}{\ 2\lambda\ }\left\{
	(x+y)^{2}+(y+x{'})^{2}\right\}+
	\frac{\ \lambda\ }{\ y\ }
	\left(\frac{\ 1\ }{\ x\ }+\frac{\ 1\ }{\ x{'}\ }\right)\right]\\
	+&
	\exp\left[-\frac{1}{\ 2\lambda\ }\left\{
	(x-y)^{2}+(y+x{'})^{2}\right\}-
	\frac{\ \lambda\ }{\ y\ }
	\left(\frac{\ 1\ }{\ x\ }-\frac{\ 1\ }{\ x{'}\ }\right)\right]\\
	+&
	\exp\left[-\frac{1}{\ 2\lambda\ }\left\{
	(x+y)^{2}+(y-x{'})^{2}\right\}+
	\frac{\ \lambda\ }{\ y\ }
	\left(\frac{\ 1\ }{\ x\ }-\frac{\ 1\ }{\ x{'}\ }\right)\right]
\end{aligned}
\end{equation}
with an overall factor $1/(2\pi\lambda)$. If we replace $y$ by $-y$ in the second term, we obtain the same expression as the first term so that we can extend the domain of the integration with respect to $y$ to the whole real axis.
The same happens in the fourth term. We thus find that we should integrate the first and the third terms along the real line. For infinitesimally small $\epsilon$, it will be evaluated by the saddle point method. For the first term, the integration will be dominated by contributions around $y=(x+x{'})/2$ while the saddle point being located at $y=(x-x{'})/2$ for the third term. By taking these contributions into account, we obtain
\begin{equation}
\label{eq:reproduce2}
\begin{aligned}
	\int_{0}^{\infty}\!\!K(x,y;\epsilon)K(y,x{'};\epsilon)\,dy
	\sim&
	\frac{1}{\ \sqrt{4\pi\lambda\,}\ }\left[
	\exp\left\{-\frac{1}{\ 4\lambda\ }(x-x{'})^{2}-
	\frac{\ 2\lambda\ }{\ xx{'}\ }\right\}\right.\\
	&\hphantom{\frac{1}{\ \sqrt{4\pi\lambda\,}\ }}\!\!\!
	\left.+
	\exp\left\{-\frac{1}{\ 4\lambda\ }(x+x{'})^{2}+
	\frac{\ 2\lambda\ }{\ xx{'}\ }\right\}\right]\\
	=&K(x,x{'};2\epsilon).
\end{aligned}
\end{equation}
Interestingly, this is again the exact result though the calculation is based on the saddle point approximation. On the four terms in \eqref{eq:product}, the first term is the product of contributions by the direct path, the second term is the one of contributions by the reflected path, and the third and the fourth terms are the cross terms of them. As we have seen above, the product of contributions from the reflected path switches into a half of the component for the direct path in the resulting kernel and the reflected component has been generated by the cross terms.

We have to add here the important remark; all observations made above are justified only for infinitesimally small $\epsilon$ becase decomposition of the short time kernel $K(x,x{'};\epsilon)$ given by \eqref{eq:feynman04} is obtained after the exponentiation of terms proportional to $\lambda$ in
\begin{equation}
\label{eq:feynman04a}
\begin{aligned}
	K(x,y;\epsilon)=&\frac{1}{\ \sqrt{2\pi\lambda\,}\ }\left[
	\left(1-\frac{\ \lambda\ }{\ xx{'}\ }\right)
	\exp\left\{-\frac{1}{\ 2\lambda\ }(x-x{'})^{2}\right\}\right.\\
	&\left.\hphantom{\frac{1}{\ \sqrt{2\pi\lambda\,}\ }}\!\!\!\!\!+
	\left(1+\frac{\ \lambda\ }{\ xx{'}\ }\right)
	\exp\left\{-\frac{1}{\ 2\lambda\ }(x+x{'})^{2}\right\}\right].
\end{aligned}
\end{equation}
 It is clear that we cannot set $x=0$ in each term in the decomposition above separately. The boundary condition, given by $K(0,x{'};\epsilon)=0$, is the result of cancellation in $\cosh z-z^{-1}\sinh z$ at $z=0$. Therefore our observation on the vanishing property of the short time kernel at the origin is valid only in the form given by \eqref{eq:feynman04}. Furthermore, the multiplication of two kernels cannot be evaluated if we write the kernels in the form of \eqref{eq:feynman04a} because we meet divergent integrals in
\begin{equation}
\label{eq:reproduce2a}
	\frac{1}{\ 2\pi\lambda\ }\int_{-\infty}^{\infty}\!\!
	\left(1-\frac{\lambda}{\ xy\ }\right)
	\left(1-\frac{\lambda}{\ x{'}y\ }\right)
	\exp\left[-\frac{1}{\ 2\lambda\ }
	\left\{(x-y)^{2}+(y-x{'})^{2}\right\}\right]\,dy
\end{equation}
in which the integrand has been obtained from the product of components of the direct path and the one of contributions of the reflected path in each kernel.
To carry out the integration of $K(x,y;\epsilon)K(y,x{'};\epsilon)$ with respect to $y$, we should not write the kernel in the form given by \eqref{eq:feynman04a}; we should rather retain $\cosh z$ and $z^{-1}\sinh z$ for $z=xy/\lambda$ and $x{'}y/\lambda$. Then, by knowing that $I(a,b)$ defined by
\begin{equation}
\label{reproduce2b}
	I(a,b)=\int_{0}^{\infty}\!\!
	e^{-x^{2}/(4K^{2})}
	\cosh\left(\frac{\ ax\ }{\lambda_{1}}\right)
	\cosh\left(\frac{\ bx\ }{\lambda_{2}}\right)\,dx,\quad
	K=\sqrt{\frac{\lambda_{1}\lambda_{2}}{\ 2(\lambda_{1}+\lambda_{2})\ }\,},
\end{equation}
is found to be
\begin{equation}
\label{eq:reproduce2c}
	I(a,b)=\frac{\ \sqrt{\pi\,}K\ }{2}\left[
	\exp\left\{K^{2}\left(\frac{\ a\ }{\ \lambda_{1}\ }+
	\frac{\ b\ }{\ \lambda_{2}\ }\right)^{2}\right\}+
	\exp\left\{K^{2}\left(\frac{\ a\ }{\ \lambda_{1}\ }-
	\frac{\ b\ }{\ \lambda_{2}\ }\right)^{2}\right\}\right],
\end{equation}
we calculate, for example,
\begin{equation}
\label{eq:reproduce2d}
	\int_{0}^{\infty}\!\!
	e^{-x^{2}/(4K^{2})}
	\frac{\lambda_{1}}{\ ax\ }\sinh\left(\frac{\ ax\ }{\lambda_{1}}\right)
	\cosh\left(\frac{\ bx\ }{\lambda_{2}}\right)\,dx=
	\frac{\ 1\ }{\ a\ }\int_{0}^{a}\!\!
	I(\alpha,b)\,d\alpha
\end{equation}
to find
\begin{equation}
\label{eq:reproduce2e}
\begin{aligned}
	&\int_{0}^{\infty}\!\!
	e^{-x^{2}/(4K^{2})}
	\frac{\lambda_{1}}{\ ax\ }\sinh\left(\frac{\ ax\ }{\lambda_{1}}\right)
	\cosh\left(\frac{\ bx\ }{\lambda_{2}}\right)\,dx\\
	=&
	-\frac{\ \pi i\lambda_{1}\ }{\ 4a\ }\left\{
	\erf\left(iK\left(\frac{\ a\ }{\ \lambda_{1}\ }+
	\frac{\ b\ }{\ \lambda_{2}\ }\right)\right)+
	\erf\left(iK\left(\frac{\ a\ }{\ \lambda_{1}\ }-
	\frac{\ b\ }{\ \lambda_{2}\ }\right)\right)\right\},
\end{aligned}
\end{equation}
where the error function is defined by
\begin{equation}
\label{eq:error_funct}
	\erf(x)=\frac{\ 2\ }{\ \sqrt{\pi\,}\ }\int_{0}^{x}\!\!
	e^{-t^{2}}\,dt.
\end{equation}
In the same way, we obtain
\begin{equation}
\label{eq:reproduce2f}
\begin{aligned}
	&\int_{0}^{\infty}\!\!
	e^{-x^{2}/(4K^{2})}
	\frac{\lambda_{2}}{\ ax\ }\sinh\left(\frac{\ bx\ }{\lambda_{2}}\right)
	\cosh\left(\frac{\ ax\ }{\lambda_{1}}\right)\,dx\\
	=&
	-\frac{\ \pi i\lambda_{2}\ }{\ 4b\ }\left\{
	\erf\left(iK\left(\frac{\ a\ }{\ \lambda_{1}\ }+
	\frac{\ b\ }{\ \lambda_{2}\ }\right)\right)-
	\erf\left(iK\left(\frac{\ a\ }{\ \lambda_{1}\ }-
	\frac{\ b\ }{\ \lambda_{2}\ }\right)\right)\right\}
\end{aligned}
\end{equation}
and
\begin{equation}
\label{eq:reproduce2g}
\begin{aligned}
	&\int_{0}^{\infty}\!\!
	e^{-x^{2}/(4K^{2})}
	\frac{\lambda_{1}\lambda_{2}}{\ abx^{2}\ }
	\sinh\left(\frac{\ ax\ }{\lambda_{1}}\right)
	\sinh\left(\frac{\ bx\ }{\lambda_{2}}\right)\,dx\\
	=&
	-\frac{\ \pi i\lambda_{1}\lambda_{2}\ }{\ 4ab\ }\left[
	\left(\frac{\ a\ }{\ \lambda_{1}\ }+
	\frac{\ b\ }{\ \lambda_{2}\ }\right)
	\erf\left(iK\left(\frac{\ a\ }{\ \lambda_{1}\ }+
	\frac{\ b\ }{\ \lambda_{2}\ }\right)\right)\right.\\
	&-
	\left(\frac{\ a\ }{\ \lambda_{1}\ }-
	\frac{\ b\ }{\ \lambda_{2}\ }\right)
	\erf\left(iK\left(\frac{\ a\ }{\ \lambda_{1}\ }-
	\frac{\ b\ }{\ \lambda_{2}\ }\right)\right)\\
	&\left.+
	\frac{i}{\ K\sqrt{\pi\,}\ }\left\{
	\exp\left\{K^{2}\left(\frac{\ a\ }{\ \lambda_{1}\ }+
	\frac{\ b\ }{\ \lambda_{2}\ }\right)^{2}\right\}-
	\exp\left\{K^{2}\left(\frac{\ a\ }{\ \lambda_{1}\ }-
	\frac{\ b\ }{\ \lambda_{2}\ }\right)^{2}\right\}
	\right\}\right],
\end{aligned}
\end{equation}
where use has been made of an indefinite integral
\begin{equation}
\label{eq:error_int}
	\int\!\!\erf(ax)\,dx=\frac{\ e^{-a^{2}x^{2}}\ }{\ a\sqrt{\pi\,}\ }+
	x\erf(ax). 
\end{equation}
Since we can express, by writing $\hbar\epsilon_{1,2}/m$ as $\lambda_{1,2}$,
\begin{equation}
\label{eq:reproduce2h}
\begin{aligned}
	&\int_{0}^{\infty}\!\!
	K(a,x;\epsilon_{1})K(x,b;\epsilon_{2})\,dx=
	\frac{1}{\ 2\pi\sqrt{\lambda_{1}\lambda_{2}\,}\ }
	\exp\left(-\frac{a^{2}}{\ 2\lambda_{1}\ }-
	\frac{b^{2}}{\ 2\lambda_{2}\ }\right)\\
	&\times\left\{
	I(a,b)-\frac{\ 1\ }{\ a\ }\int_{0}^{a}\!\!I(\alpha,b)\,d\alpha
	-\frac{\ 1\ }{\ b\ }\int_{0}^{b}\!\!I(a,\beta)\,d\beta+
	\frac{\ 1\ }{\ ab\ }\int_{0}^{a}\int_{0}^{b}\!\!I(\alpha,\beta)
	\,d\alpha\,d\beta\right\},
\end{aligned}
\end{equation}
we find
\begin{multline}
\label{eq:reproduce2i}
	\int_{0}^{\infty}\!\!
	K(a,x;\epsilon_{1})K(x,b;\epsilon_{2})\,dx=
	\frac{1}{\ \sqrt{2\pi(\lambda_{1}+\lambda_{2})\,}\ }
	\exp\left(-\frac{a^{2}}{\ 2\lambda_{1}\ }-
	\frac{b^{2}}{\ 2\lambda_{2}\ }\right)\\
	\times\left[
	\left(1+\frac{\ \lambda_{1}+\lambda_{2}\ }{ab}\right)
	\exp\left\{K^{2}\left(\frac{\ a\ }{\ \lambda_{1}\ }+
	\frac{\ b\ }{\ \lambda_{2}\ }\right)^{2}\right\}\right.\\
	\left.+
	\left(1-\frac{\ \lambda_{1}+\lambda_{2}\ }{ab}\right)
	\exp\left\{K^{2}\left(\frac{\ a\ }{\ \lambda_{1}\ }-
	\frac{\ b\ }{\ \lambda_{2}\ }\right)^{2}\right\}
	\right]
\end{multline}
which can be immediately rewritten as 
\begin{equation}
\label{eq:reproduce2j}
\begin{aligned}
	\int_{0}^{\infty}\!\!
	&K(a,x;\epsilon_{1})K(x,b;\epsilon_{2})\,dx
	=
	\sqrt{\frac{2}{\ \pi(\lambda_{1}+\lambda_{2})\ }\,}\\
	&\times
	\exp\left\{-\frac{\ a^{2}+b^{2}\ }{\ 2(\lambda_{1}+\lambda_{2})\ }\right\}
	%&\hphantom{\int_{0}^{\infty}\!\!
	%K(a,x;\epsilon_{1})}\times
	\left\{\cosh\left(\frac{ab}{\ \lambda_{1}+\lambda_{2}\ }\right)-
	\frac{\ \lambda_{1}+\lambda_{2}\ }{ab}
	\sinh\left(\frac{ab}{\ \lambda_{1}+\lambda_{2}\ }\right)\right\}\\
	=&K(a,b,\epsilon_{1}+\epsilon_{2}).
\end{aligned}
\end{equation}
We have thus confirmed the reproducing property of the Feynman kernel in a rigorous manner. We may expect that the use of the technique shown above even for the divergent integrals in \eqref{eq:reproduce2a} and its counter part from the products of contributions by direct and reflected paths will make these integrals finite. Such a mathematically incorrect method is not admissible though interesting and convenient. Calculation given above to clarify the reproducing property of the Feynman kernel clearly shows that the idea to decompose the Feynman kernel into a sum over paths for a finite imaginary time $\beta$ is of no use and that it has its meaning only when we restrict ourselves to an infinitesimally small $\epsilon$.

It will be hard to evaluate the time sliced path integral defined by \eqref{eq:feynman01} for generic values of $\nu$ under the existence of a potential $V(x)$. The radial oscillator is the exception for which we can calculate the Feynman kernel exactly to result in\cite{KleinertBook,Sakoda2017}
\begin{equation}
\label{eq:feynman05}
	K(a,b;\beta)=
	\frac{\ m\omega \sqrt{ab\,}\ }{\ \hbar\sinh\omega\beta\ }\exp\left[
	-\frac{\ m\omega\coth\omega\beta\ }{\ 2\hbar\ }
	(a^{2}+b^{2})\right]
	I_{\nu-1/2}\left(\frac{\ m\omega ab\ }{\ \hbar\sinh\omega\beta\ }\right).
\end{equation}
Our formulation in this article is also applicable for constructing the Duru-Kleinert(DK) path integral in a pseudotime sliced form\cite{Sakoda2017}. We can then prove the DK-equivalence of path integrals thus formulated for the radial Coulomb system and that of the radial oscillator because they both result in the path integral of a system with the Morse potential.

We shall finally try to consider decomposition of the short time kernel \eqref{eq:kernel05} for an arbitrary potential $V(x)$ into a sum over paths.
The contribution from the direct path that connects $x$ to $x{'}$ directly in the short time interval $\epsilon$ is given by
\begin{equation}
\label{eq:kernel06}
	\frac{1}{\ \sqrt{2\pi\hbar\epsilon\,}\ }
	\exp\left\{-\frac{1}{\ 2\lambda\ }(x-x{'})^{2}-
	\frac{\ \nu(\nu-1)\lambda\ }{2xx{'}}-
	\frac{\ \epsilon\ }{\ \hbar\ }V(\sqrt{xx{'}\,})\right\}
\end{equation}
by keeping $k=0$ and $k=1$ terms in the first sum in \eqref{eq:asymp01}.
To find the contribution from the reflected path, we must first replace $\sqrt{xx{'}\,}$ by $\sqrt{-xx{'}\,}$ then we take the second term in \eqref{eq:asymp01}. In other words, we set $\arg(xx{'})=\pi$ to write $xx{'}=e^{\pi i}\abs{xx{'}}$ in the modified Bessel function then we take the first term in \eqref{eq:asymp01} for $e^{\nu\pi i}\sqrt{\abs{xx{'}}\,}I_{\nu-1/2}(m\abs{xx{'}}/(\hbar\epsilon))$. We thus obtain
\begin{equation}
\label{eq:kernel07}
	\frac{e^{\nu\pi i}}{\ \sqrt{2\pi\lambda\,}\ }
	\exp\left\{-\frac{1}{\ 2\lambda\ }(x+x{'})^{2}
	-\frac{\ \nu(\nu-1)\lambda\ }{2xx{'}}-
	\frac{\ \epsilon\ }{\ \hbar\ }V(\sqrt{-xx{'}\,})\right\}
\end{equation}
as the contribution from the reflected path. Therefore the short time kernel, given by \eqref{eq:kernel05}, is now rewritten as
\begin{equation}
\label{eq:kernel08}
\begin{aligned}
	&K(x,x{'};\epsilon)\\
	=&
	\frac{1}{\ \sqrt{2\pi\hbar\epsilon\,}\ }
	\exp\left\{-\frac{1}{\ 2\lambda\ }(x-x{'})^{2}-
	\frac{\ \nu(\nu-1)\lambda\ }{2xx{'}}-
	\frac{\ \epsilon\ }{\ \hbar\ }V(\sqrt{xx{'}\,})\right\}+\\
	&
	\frac{e^{\nu\pi i}}{\ \sqrt{2\pi\lambda\,}\ }
	\exp\left\{-\frac{1}{\ 2\lambda\ }(x+x{'})^{2}-
	\frac{\ \nu(\nu-1)\lambda\ }{2xx{'}}-
	\frac{\ \epsilon\ }{\ \hbar\ }V(\sqrt{-xx{'}\,})\right\}.
\end{aligned}
\end{equation}
We may equally set $\arg(xx{'})=-\pi$ for the contribution by the reflected path. The phase factor for this choice is then given by $e^{-\nu\pi i}$ instead of $e^{\nu\pi i}$ above. It is possible to calculate an integration of a product $K(x,y;\epsilon)K(y,x{'};\epsilon)$ with respect to $y$ in a similar manner as we have done to obtain \eqref{eq:reproduce2}. In doing so, we set $\arg(x)=-\pi$ in the contribution of the reflected path for $K(x,y;\epsilon)$ and $\arg(x{'})=\pi$ in the same part of $K(y,x{'};\epsilon)$. Then the product of contributions from direct path and the one of contributions from the reflected path are combined to generate the contribution from the direct path in the resulting kernel as before. The cross terms of these contributions yield
\begin{equation}
\label{eq:reproduce3}
\begin{aligned}
	&\frac{m}{\ 2\pi\hbar\epsilon\ }e^{-\nu\pi i}
	\int_{0}^{\infty}\!\!dy\,
	\exp\left[-\frac{m}{\ 2\hbar\epsilon\ }\left\{
	(x+y)^{2}+(y-x{'})^{2}\right\}\right]\\
	&\times\exp\left[
	\frac{\ \nu(\nu-1)\hbar\epsilon\ }{2my}
	\left(\frac{\ 1\ }{\ x\ }-\frac{\ 1\ }{\ x{'}\ }\right)-
	\frac{\ \epsilon\ }{\ \hbar\ }
	\left\{V(\sqrt{-xy\,})+V(\sqrt{x{'}y\,})\right\}
	\right]+\\
	&
	\frac{m}{\ 2\pi\hbar\epsilon\ }e^{\nu\pi i}
	\int_{0}^{\infty}\!\!dy\,
	\exp\left[-\frac{m}{\ 2\hbar\epsilon\ }\left\{
	(x-y)^{2}+(y+x{'})^{2}\right\}\right]\\
	&\times\exp\left[
	-\frac{\ \nu(\nu-1)\hbar\epsilon\ }{2my}
	\left(\frac{\ 1\ }{\ x\ }-\frac{\ 1\ }{\ x{'}\ }\right)-
	\frac{\ \epsilon\ }{\ \hbar\ }
	\left\{V(\sqrt{xy\,})+V(\sqrt{-x{'}y\,})\right\}
	\right].
\end{aligned}
\end{equation}
In the first integral, we may replace $y$ by $-y=e^{\pi i}y$ to obtain the same integrand as the one in the second integral. Since this integrand has been generated from a product of modified Bessel functions, the replacement of $y$ by $e^{\pi i}y$ produces a phase factor $e^{2\nu\pi i}$. Then we obtain
\begin{equation}
\label{eq:reproduce4}
\begin{aligned}
	&\frac{m}{\ 2\pi\hbar\epsilon\ }e^{\nu\pi i}
	\int_{-\infty}^{\infty}\!\!dy\,
	\exp\left[-\frac{m}{\ 2\hbar\epsilon\ }\left\{
	(x-y)^{2}+(y+x{'})^{2}\right\}\right]\\
	&\times\exp\left[
	-\frac{\ \nu(\nu-1)\hbar\epsilon\ }{2my}
	\left(\frac{\ 1\ }{\ x\ }-\frac{\ 1\ }{\ x{'}\ }\right)-
	\frac{\ \epsilon\ }{\ \hbar\ }
	\left\{V(\sqrt{xy\,})+V(\sqrt{-x{'}y\,})\right\}
	\right]\\
	\sim&
	e^{\nu\pi i}\sqrt{\frac{m}{\ 4\pi\hbar\epsilon\ }\,}
	\exp\left[-\frac{m}{\ 4\hbar\epsilon\ }(x+x{'})^{2}-
	\frac{\ 2\epsilon\ }{\ \hbar\ }\left\{
	-\frac{\ \nu(\nu-1)\hbar^{2}\ }{2mxx{'}}+V(\sqrt{-xx{'}\,})\right\}\right]
\end{aligned}
\end{equation}
as a result of the saddle point approximation. We thus observe again that the cross terms of contributions from the direct path and the reflected path in a product of two short time kernels are combined to produce the contribution by the reflected path in the resulting kernel though our calculation will be acceptable only for infinitesimally small $\epsilon$. It should be noted here that decomposition of the short time kernel into sum over paths, given by \eqref{eq:kernel08}, is exact if we set $\nu=1$ even when a potential $V(x)$ exists. In this regard, the free particle and the system for $\nu=1$ with a potential are quite exceptional.

\section{Conclusion}
We have developed a method to construct the time sliced path integral for the Feynman kernel of a particle on the half-line by making use of the eigenfunction for the Hamiltonian whose eigenfunction being expressed in terms of the Bessel function. The short time kernel thus obtained can be viewed as the short time version of the Hamiltonian path integral and can be converted into the Lagrangian form. It is the second sum in the asymptotic expansion of the modified Bessel function that picks up the contribution from the reflected path  to the Feynman kernel for a particle on the half-line. 
The contribution from the reflected path is accompanied with a phase factor that depends on the parameter of the potential. If the parameter is set to be an odd integer, the phase factor reduces to $-1$ and the contribution from the reflected path is subtracted from the one by the direct path. For other values of the parameter, however, this behavior is changed and becomes additive contribution if we set the parameter to be an even integer. We therefore conclude that the factor $-1$ for the free particle cannot be universal and the phase by the reflection depends on the dynamics of the system.
We have also confirmed that the contribution from the reflected path propagates through the cross terms of contributions from the direct and the reflected paths and also that the product of contributions from reflected paths is combined with the one of contributions from direct paths to result in the contribution from the direct path in the reproduced kernel. Decomposition into a sum over paths for the Feynman kernel is, however, restricted to the short time kernel for an infinitesimally small imaginary time. The free particle or the system for $\nu=1$ with a potential $V(x)$ are exceptions in this regard because for these systems we can always write the kernel as a sum of contributions from direct and reflected paths.

\end{document}